\documentstyle[11pt,twoside,sympo2011,epsfig]{article}
\begin{document}
\title{The NStED Periodogram Service \& Interface for Public CoRoT Data}
\author{K. von Braun, M. Abajian, A. Beekley, G. B. Berriman, G. Bryden, B. Chan, D. R. Ciardi, J. Good, M. Harbut, S. R. Kane, A. Laity, C. Lau, M. Lynn, D. McElroy, P. Plavchan, M. Regelson, R. Rey, S. V. Ramirez, J. Stauffer, \& A. Zhang}
\affil{NASA Exoplanet Science Institute, Infrared Processing and Analysis Center, California Institute of Technology [kaspar@caltech.edu]} 
\begin{abstract}
As part of the NASA-CNES agreement, the NASA Star and Exoplanet Database (NStED) serves as the official US portal for the public CoRoT data products. NStED is a general purpose archive with the aim of providing support for NASA's planet finding and characterization goals. Consequently, the NASA Exoplanet Science Institute (NExScI) developed, and NStED adapted, a periodogram service for CoRoT data to determine periods of variability phenomena and create phased photometric light curves. Through the NStED periodogram interface, the user may choose three different period detection algorithms to use on any photometric time series product, or even upload and analyze their own data. Additionally, the NStED periodogram is remotely accessed by the CoRoT archive as part of its interface. NStED is available at {\bf http://nsted.ipac.caltech.edu}.
\end{abstract}
\section{Introduction}

NStED's periodogram service provides a flexible user interface to search for periodic signals 
in the public CoRoT asteroseismology and exoplanet data, and to return their downloadable, phased light curves. In particular, it features the following:

\begin{itemize}

\item Convenience: the periodogram service is available for all public CoRoT and Kepler data. In addition, it can be applied to user uploaded light curves. 

\item Flexibility: Returns periodograms and phased light curves and computes phased light curves for user-supplied periods, thereby supporting three common algorithms\footnote{See http://nsted.ipac.caltech.edu/periodogram/applications/Periodogram/docs/Algorithms.html.}.

\item Speed: It is parallelized across a 128-core cluster and can process a 100,000-point light curve in less than a minute. 

\item Customization: Allows user to adjust parameters and select new algorithms from return pages. 

\end{itemize}

\noindent Figure 1 illustrates an example operation of the periodogram service on the public exoplanet data for CoRoT-1. The user starts with the raw light curve (top panel), available from NStED. The periodogram service calculates the power contained in a range of periods (middle panel) and allows the user to phase by any period corresponding a peak in the spectrum (or indeed any user defined period). The selection of the period with the highest peak returns the phased light curve (bottom panel), which is downloadable by the user. More information is available at: {\bf http://www.nsted.ipac.caltech.edu/Periodogram}.

%
%


\begin{figure}[h]
\begin{center}
\epsfig{width=10cm,file=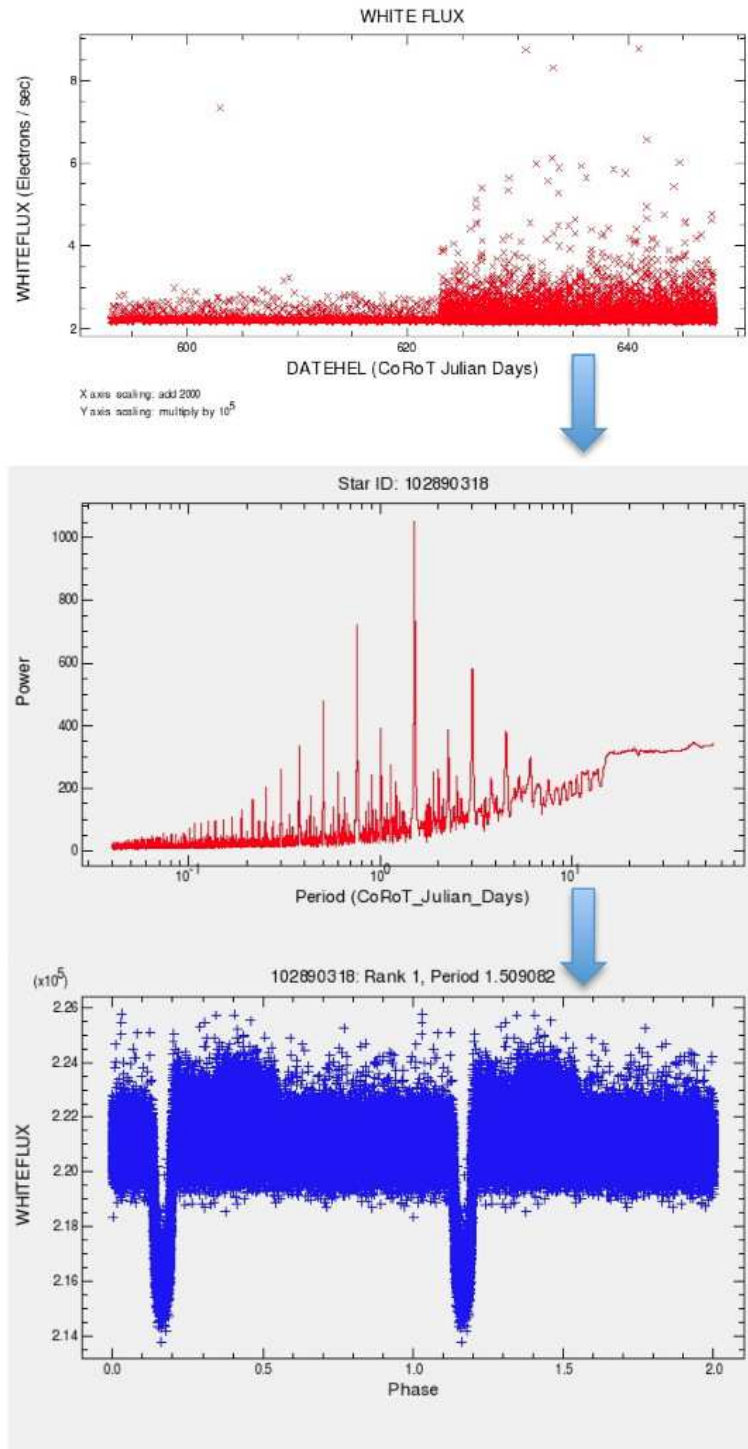}
\caption{Illustration of NStED's Periodogram Service. {\bf Top Panel:} The raw light curve of CoRoT-1. {\bf Middle Panel:} The NStED periodgram, indicating the amount of power for a range of periods. {\bf Bottom Panel:} CoRoT-1's light curve, phased by the period with the highest peak in the periodgram.}
\end{center}
\end{figure}





\end{document}